\documentclass[aps,pra,twocolumn,showpacs,superscriptaddress]{revtex4} 
\usepackage{graphicx}
\usepackage{amsmath}
\usepackage{amssymb}
\usepackage{epstopdf}
\usepackage{amsfonts}
\usepackage{dcolumn}
\begin{document}
\title{One-dimensional Fermi gas with a single impurity
in a harmonic trap: Perturbative
description of the upper branch}
\author{Seyed Ebrahim Gharashi, X. Y. Yin, Yangqian Yan, and D. Blume}
\address{Department of Physics and Astronomy,
Washington State University,
  Pullman, Washington 99164-2814, USA}
\date{\today}

\begin{abstract}
The transition from ``few to many'' has 
recently been probed experimentally in an ultracold harmonically
confined 
one-dimensional
lithium gas, in which a single impurity atom
interacts with a background gas consisting of one, two,
or more identical fermions
[A. N. Wenz {\em{et al.}}, Science {\bf{342}}, 457 (2013)].
For repulsive interactions between the 
background or majority atoms and the
impurity, the interaction energy 
for relatively moderate system sizes
was analyzed and found to 
converge toward the corresponding expression
for an infinitely
large Fermi gas. 
Motivated by these experimental results, we apply perturbative techniques
to determine the interaction energy
for weak and strong
coupling 
strengths
and derive approximate descriptions
for the interaction energy for repulsive interactions
with varying strength between
the impurity and the majority atoms and any number of
majority atoms.

\end{abstract}
\maketitle

\section{Introduction}
One-dimensional Bose and Fermi systems with contact interactions have been
studied for many decades now, especially
in the regime where the systems obey periodic boundary
conditions~\cite{1Dbook,giamarchi,RMP1,RMP2,1Dbook2}. A large fraction of the
eigenstates can be thought of as 
corresponding to gas-like states.
A second subset of eigenstates
corresponds to
self-bound droplet-like states.
These states maintain their bound state character
in the absence of periodic boundary conditions, i.e., in free space.
In many cases,
both the gas-like and 
droplet-like states can be obtained analytically via the Bethe ansatz.
The Bethe ansatz takes advantage of the fact that 
the zero-range nature of the interactions, 
combined with the fact that 
particles in one dimension have to pass through each other to 
exchange positions, allows one to identify constants of motion.
The solutions can then be derived in terms of these constants of motion.
A closely related aspect is that 
a variety of
one-dimensional systems 
with two-body contact interactions
are
integrable~\cite{1Dbook,RMP2}.

The solution of the homogeneous system can be applied to one-dimensional
systems under spatially varying external confinement
via the local density 
approximation~\cite{menotti,astrakharchik,oldPRLwithPitaevskii,tokatly,orso}.
This approximation typically provides a highly accurate description
for a large number of particles
but not necessarily for a small number of particles.
It is thus desirable to derive more accurate descriptions
for small one-dimensional systems with two-body
delta-function interactions under
external confinement. Unfortunately,
extensions of the Bethe ansatz 
to inhomogeneous systems
are, in general, not known.
This can be understood 
intuitively by realizing that the relative two-body momentum in 
inhomogeneous
systems is not
conserved due to the presence of the spatially varying confinement.
Correspondingly, harmonically trapped one-dimensional few-body
systems have been
treated numerically by various 
techniques~\cite{astrakharchik,mueller,brouzos,harshman,Gharashi12,Gharashi13,conduit,lewenstein}.

In this work, we 
apply standard Raleigh-Schr\"odinger perturbation theory 
to harmonically confined systems and derive approximate
solutions whose accuracy can be improved systematically 
by considering successively 
higher orders
in the 
expansion 
in the small
parameter.
We focus on one-dimensional Fermi gases with a 
single impurity under external harmonic confinement.
This system is of particular interest since it has been realized
experimentally in Jochim's cold atom 
laboratory~\cite{Jochim1,Jochim3}.
In the experiments, the impurity is a lithium 
atom that occupies
a
hyperfine state different from the hyperfine 
state that the majority atoms occupy.
The trapping geometry is highly-elongated and 
effectively
one-dimensional.
We will show that our perturbative results enable us to 
calculate the energy of the upper branch, which has been
studied experimentally, with fairly
good accuracy for all $N$ over a wide range of coupling strengths.
In addition, our results provide bounds on the energies in the
weakly- and strongly-interacting regimes. These bounds can, e.g., be used
to assess the accuracy of numerical solutions.

The remainder of this paper is organized as follows.
Section~\ref{sec_system}
introduces the system Hamiltonian and notation.
Section~\ref{sec_pert} summarizes our perturbative results.
The perturbative results are analyzed in Secs.~\ref{sec_fitting}
and \ref{sec_implication}.
Finally, Sec.~\ref{sec_conclusion} concludes.

\section{System Hamiltonian}
\label{sec_system}
We consider a single impurity immersed in
a one-dimensional Fermi gas that consists of $N$ identical 
mass $m$ fermions. The mass
of the impurity is equal to that of the majority or background 
particles.
The impurity, with position coordinate $z_0$, interacts with the  
majority particles, with position coordinates
$z_j$ ($j=1,\cdots,N$), through a
zero-range two-body potential with strength
$g$,
\begin{eqnarray}
V_{2\text{b}}(z_{j0}) = g 
\delta(z_j-z_0),
\label{eq_vtwobody}
\end{eqnarray}
where $z_{j0}=z_j-z_0$.
The Hamiltonian $H$ for the harmonically confined $(N,1)$ system
then reads
\begin{eqnarray}
\label{eq_ham}
H = 
\sum_{j=1}^N H_{\text{ho}}(z_j) +
H_{\text{ho}}(z_0) + 
\sum_{j=1}^N V_{2{\text{b}}}(z_{j0}),
\end{eqnarray}
where the single particle harmonic oscillator Hamiltonian
$H_{\text{ho}}(z)$ is given by
\begin{eqnarray}
H_{\text{ho}}(z)=
-\frac{\hbar^2}{2m} \frac{\partial^2}{\partial z^2} + \frac{1}{2}
m \omega^2 z^2;
\end{eqnarray}
here, $\omega$ denotes the angular trapping frequency.
The delta-function interactions in Eq.~(\ref{eq_ham}) can be replaced by
a set of boundary conditions on the many-body wave function
$\Psi(z_0,z_1,\cdots,z_N)$,
\begin{eqnarray}
\left(
\frac{\partial \Psi}{\partial z_{j0}}\bigg|_{z_{j0} \rightarrow 0^+} -
\frac{\partial \Psi}{\partial z_{j0}}\bigg|_{z_{j0} \rightarrow 0^-} 
\right )
= \frac{g m }{\hbar^2} \Psi|_{z_{j0} \rightarrow 0},
\end{eqnarray}
where the limits $z_{j0} \rightarrow 0^+$, 
$z_{j0} \rightarrow 0^-$,  and $z_{j0} \rightarrow 0$
are taken while keeping the other $N$ coordinates, i.e., 
$z_1,\cdots,z_{j-1},z_{j+1},\cdots,z_N$ and $(z_j+z_0)/2$, fixed.

\begin{figure}
\centering
\includegraphics[angle=0,width=0.35\textwidth]{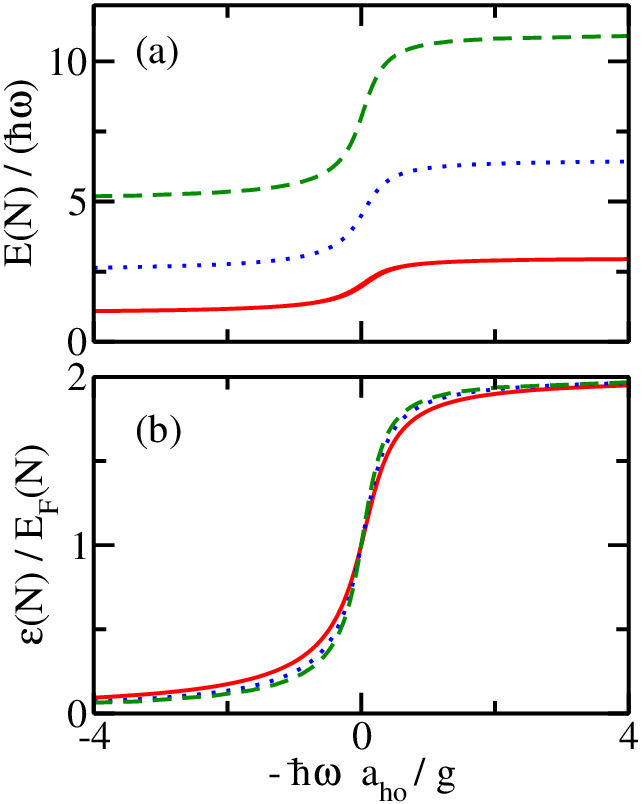}
\vspace*{0.5in}
\caption{(Color online)
(a)
Solid, dotted, and dashed lines
show the energy of the upper
branch for $N=1$, 2, and 3 as a function of $-1/g$.
The energies of the $(1,1)$ system are obtained by solving the transcendental
equation derived in Ref.~\cite{Busch98}.
The energies of the $(2,1)$ and $(3,1)$ system are taken from 
Refs.~\cite{Gharashi12,Gharashi13,GharashiNote}.
(b)
Solid, dotted, and dashed lines show the interaction energy
$\epsilon(N)$, normalized
by the Fermi energy $E_{\text{F}}(N)$,
for $N=1$, 2, and $3$, respectively, as a 
function of $-1/g$.
The harmonic oscillator length $a_{\text{ho}}$ is
defined in Eq.~(\ref{eq_aho}).}\label{fig1}
\end{figure}

In the following, we determine the eigenenergies $E(N)$ of the 
Hamiltonian $H$ for various $N$.
Throughout, we restrict ourselves to the 
so-called upper branch. This branch can be populated
by preparing the system in the non-interacting limit ($g \rightarrow 0^+$)
and by then adiabatically first increasing $g$ to large positive
values,
then continuing across the confinement-induced
resonance~\cite{olsh} to infinitely negative $g$
values and finally increasing
$g$ to small negative values.
Solid, dotted, and dashed lines in Fig.~\ref{fig1}(a)
show the energy of the upper branch for $N=1$~\cite{Busch98}, 
2, and 3~\cite{Gharashi12,Gharashi13,GharashiNote},
respectively, as a function of $-1/g$.
For all $N$, the energy increases monotonically as a 
function of increasing $-1/g$.
The upper branch corresponds
to the ground state of the model Hamiltonian
when $g$ is positive but not when $g$ is negative. For negative
$g$, the model Hamiltonian supports molecular-like bound states.
In real cold atom
systems, energetically lower-lying molecular states exist
even for positive $g$. 
However, it has been demonstrated
experimentally~\cite{Jochim3} that the upper branch 
can be populated with reasonably high fidelity
for positive $g$,
motivating us---as well as others~\cite{astrakharchik,brouzos,harshman,conduit,lewenstein,volosniev1,volosniev2,ho,levinsen}---to 
investigate the properties of the upper branch within a
stationary zero-temperature quantum mechanics framework.
Since decay to states with molecular
character can lead to significant depo\-pu\-lation of the upper branch
for negative $g$, 
our primary focus in the following lies on the positive $g$ portion
of the upper branch.

For $g=0^+$, the energy of the upper branch is equal to
$E_{\text{ni}}(N)=(N^2+1) \hbar \omega/2$.
We write the energy $E(N)$ of the upper branch
in terms of the interaction energy $\epsilon(N)$,
\begin{eqnarray}
E(N)=E_{\text{ni}}(N)+\epsilon(N).
\end{eqnarray}
Solid, dotted, and dashed lines
in Fig.~\ref{fig1}(b)
show the interaction energies,
normalized by the energy $E_{\text{F}}(N)$,
for systems with $N=1$, $2$,
and 3
majority particles. 
The energy $E_{\text{F}}(N)$ is directly proportional to the number of 
majority particles,
\begin{eqnarray}
\label{eq_fermienergy}
E_{\text{F}}(N)= N \hbar \omega.
\end{eqnarray}
Figure~\ref{fig1}(b)
shows that the normalized 
interaction energy depends relatively weakly on 
the number of particles.
Independent of $N$, we have $\epsilon(N)=0$ for $g=0^+$
and
$\epsilon(N)=E_{\text{F}}(N)$ 
for $|g|=\infty$.
As can be read off Figs.~\ref{fig1}(a) and \ref{fig1}(b),
the energy increase of the upper branch is the same on the positive
$g$ side as it is on the negative $g$ side,
indicating that $\epsilon(N)$ approaches
$2E_{\text{F}}(N)$ in the $g=0^-$ limit  for $N=1-3$.
We refer to  $E_{\text{F}}(N)$ as the Fermi energy of the majority particles.
It should be noted, however, that the ``exact'' Fermi energy
of the majority particles is $E_{\text{F}}(N)-\hbar \omega /2$, i.e.,
$E_{\text{F}}(N)$ corresponds to the leading order term of the Fermi energy
of the majority particles in the large $N$ limit. 

One of the main goals of this paper is 
to derive expansions for the interaction energy of the upper
branch around $g=0^+$ and
$|g|=\infty$ 
using standard Raleigh-Schr\"odinger perturbation
theory for any $N$, i.e., for $N=1,\cdots,\infty$.
To this end, we express the interaction energies
$\epsilon^{(0^+)}$ and
$\epsilon^{(|\infty|)}$
in the vicinity of $g=0^+$ and
$|g|=\infty$,
respectively,
in a power series of the dimensionless interaction parameter
$\gamma$ (for $|g|$ small) or
in a power series of 
$\gamma^{-1}$ (for $|g|$ large)~\cite{series},
\begin{eqnarray}
\label{eq_pertplus}
\epsilon^{(0^+)}(N)= 
\left[
\sum_{k=1}^{k_{\text{max}}} B^{(k)}(N) \gamma^k
\right] E_{\text{F}}(N) + {\cal{O}}(\gamma^{k_{\text{max}}+1})
\end{eqnarray}
and
\begin{eqnarray}
\label{eq_pertinfty}
\epsilon^{(|\infty|)}(N)= \nonumber \\
\left[1+
\sum_{k=1}^{k_{\text{max}}} C^{(k)}(N) \gamma^{-k}
\right] E_{\text{F}}(N) + {\cal{O}}(\gamma^{-(k_{\text{max}}+1)}),
\end{eqnarray}
where the dimensionless interaction parameter
$\gamma$ is given by
\begin{eqnarray}
\gamma =  
\frac{\pi}{\sqrt{2 N}} \frac{g}{\hbar \omega a_{\text{ho}}},
\label{eq_gamma}
\end{eqnarray}
with
\begin{eqnarray}
\label{eq_aho}
a_{\text{ho}} = \sqrt{\frac{\hbar }{ m \omega}}
\end{eqnarray}
denoting the harmonic oscillator
length.
As
we will show below, the scaling of the interaction energy by
$E_{\text{F}}(N)$ ensures a smooth connection between the energy shifts
for finite
and infinite $N$.
In Eqs.~(\ref{eq_pertplus})-(\ref{eq_pertinfty}),
the 
dimensionless
$k$th-order perturbation theory coefficients $B^{(k)}(N)$
and
$C^{(k)}(N)$ 
depend on $N$ and will be determined 
in the next section.

\section{Perturbative Results}
\label{sec_pert}
{\em{$N \rightarrow \infty$ limit:}}
The impurity problem for the homogeneous system
with positive $\gamma$
was solved by McGuire in 1965~\cite{mcguirepositive}.
Within the local density approximation
the Fermi wave vector is replaced by the wave vector at the trap center
such that the interaction energy
of the ground state
for the harmonically trapped system with $N \rightarrow \infty$ 
becomes~\cite{Jochim3}
\begin{eqnarray}
\label{eq_mcguire}
\frac{\epsilon(\infty)}{E_{\text{F}}(\infty)} =
\frac {\gamma}{\pi^2} 
\left[1-\frac{\gamma}{4} + 
\left(
\frac{\gamma}{2\pi} + \frac{2\pi} {\gamma} \right) 
\arctan \left(\frac{\gamma}{2\pi} 
\right) 
\right]
\label{eq_mcguire}.
\end{eqnarray}
Expanding Eq.~(\ref{eq_mcguire}) around
$\gamma = 0^+$ and $|\gamma|=\infty$, respectively, 
$B^{(k)}(\infty)$ and $C^{(k)}(\infty)$ can be obtained for $k=1,2,\cdots$.
We find
$B^{(1)}(\infty)=2/\pi^2$,
$B^{(2)}(\infty)=-1/(4\pi^2)$,
$B^{(3)}(\infty)=1/(6\pi^4)$,
$C^{(1)}(\infty)=-8/3$,
$C^{(2)}(\infty)=0$, and
$C^{(3)}(\infty)=32 \pi^2/15$.
The numerical values of these coefficients are summarized in 
Tables~\ref{tab1}-\ref{tab2}.
\begin{table}
\caption{Coefficients $B^{(k)}(N)$ 
for various $(N,k)$ combinations.
The numbers in 
parenthesis
denote the uncertainty
that arises from evaluating the perturbation theory
sums with a finite energy cutoff.
The numbers without errorbars have been rounded.
}
\label{tab1}
\centering
\begin{tabular}{r@{=}l|r@{.}l r@{.}l r@{.}l}
 \multicolumn{2}{c|}{}&  \multicolumn{2}{c}{$k=1$}  & \multicolumn{2}{c}{$k=2$} & \multicolumn{2}{c}{$k=3$} \\
\hline
$N$&$1$  & $0$&$179587$ & $-0$&$0223551$ & $0$&$00179230$ \\
$N$&$2$  & $0$&$190481$ & $-0$&$0239838$ & $0$&$00179523$ \\
$N$&$3$  & $0$&$194409$ & $-0$&$0244852$ & $0$&$00177603(1)$\\
$N$&$4$  & $0$&$196423$ & $-0$&$0247210$ & $0$&$0017627(1)$\\
$N$&$5$  & $0$&$197647$ & $-0$&$0248563$ & $0$&$0017535(1)$\\
$N$&$6$  & $0$&$198469$ & $-0$&$0249435$ & $$0&$0017470(1)$\\
$N$&$7$  & $0$&$199059$ & $-0$&$0250042$ & \multicolumn{2}{c}{}\\
$N$&$8$  & $0$&$199503$ & $-0$&$0250488$ & \multicolumn{2}{c}{}\\
$N$&$9$  & $0$&$199849$ & $-0$&$0250828$ & \multicolumn{2}{c}{}\\
$N$&$10$ & $0$&$200126$ & $-0$&$0251096$ & \multicolumn{2}{c}{}\\
$N$&$11$ & $0$&$200353$ & $-0$&$0251313$ & \multicolumn{2}{c}{}\\
$N$&$12$ & $0$&$200543$ & $-0$&$0251491$ & \multicolumn{2}{c}{}\\
$N$&$\infty$ & $0$&$202642$ & $-0$&$0253303$ & $0$&$00171100$ 
\end{tabular}
\end{table}
It is readily shown that the 
small and large $\gamma$ series, Eqs.~(\ref{eq_pertplus})
and (\ref{eq_pertinfty}),
converge for $\gamma < 2 \pi$ and $\gamma^{-1} < (2 \pi)^{-1}$, respectively.
Table~\ref{tab2} shows that $C^{(2)}(\infty)$ vanishes.
We will return to 
this
finding 
\begin{table}
\caption{Coefficients $C^{(k)}(N)$ 
for various $(N,k)$ combinations.
The numbers in 
parenthesis
denote the uncertainty
that arises from evaluating the perturbation theory
sums with a finite energy cutoff.
The numbers without errorbars have been rounded.
}
\label{tab2}
\centering
\begin{tabular}{r@{=}l|r@{.}l r@{.}l r@{.}l }
\multicolumn{2}{c|}{} &  \multicolumn{2}{c}{$k=1$}  & \multicolumn{2}{c}{$k=2$} & \multicolumn{2}{c}{$k=3$} \\
\hline
$N$&$1$ & $-3$&$54491$ & $3$&$85603$ & $34$&$3007$ \\
$N$&$2$ & $-3$&$17245$ & $2$&$41904(1)$ & $25$&$38(2)$ \\
$N$&$3$ & $-3$&$02854$ & $1$&$8142(2)$ & $23$&$78(8)$ \\
$N$&$4$ & $-2$&$95040$ & \multicolumn{2}{c}{} & \multicolumn{2}{c}{}\\
$N$&$5$ & $-2$&$90081$ & \multicolumn{2}{c}{} & \multicolumn{2}{c}{}\\
$N$&$6$ & $-2$&$86634$ & \multicolumn{2}{c}{} & \multicolumn{2}{c}{}\\
$N$&$7$ & $-2$&$84091$ & \multicolumn{2}{c}{} & \multicolumn{2}{c}{}\\
$N$&$8$ & $-2$&$82133$ & \multicolumn{2}{c}{} & \multicolumn{2}{c}{}\\
$N$&$9$ & $-2$&$80578$ & \multicolumn{2}{c}{} & \multicolumn{2}{c}{}\\
$N$&$\infty$ & $-2$&$66667$ & \multicolumn{2}{c}{$0$} & $21$&$0552$ 
\end{tabular}
\end{table}
when we discuss the $N$-dependence of the $C^{(k)}(N)$ coefficients.

{\em{$(1,1)$ system:}}
The eigenenergies of the harmonically trapped $(1,1)$ system 
can be obtained for any $\gamma$ 
by solving a simple transcendental 
equation~\cite{Busch98}.
Expanding the transcendental equation around the known
eigenenergies for small and large $\gamma$,
one obtains
power series in the interaction energy. 
Inverting these series, one obtains
analytical expressions for the $B^{(k)}(1)$ and
$C^{(k)}(1)$
coefficients. 
We find
$B^{(1)}(1)=\pi^{-3/2}$,
$B^{(2)}(1)=-\ln(2)/ \pi^{3}$,
$B^{(3)}(1)=-[\pi^2-9 \ln(4)^2]/(24 \pi^{9/2})$,
$C^{(1)}(1)=-2 \pi^{1/2}$,
$C^{(2)}(1)=-4 \pi [\ln(2)-1]$,
and
$C^{(3)}(1)=\pi^{3/2}[\pi^2-24-9(\ln(4)-4)\ln(4)]/3$.
The numerical values of these coefficients are
summarized in Tables~\ref{tab1}-\ref{tab2}.
As in the $N \rightarrow \infty$
case, the small and large
$\gamma$ series
for $N=1$, Eqs.~(\ref{eq_pertplus}) and (\ref{eq_pertinfty}),
have a finite radius of convergence.
Employing the techniques of Ref.~\cite{zama05},
we find---using up to 50
expansion coefficients---that the small and large
$\gamma$ series converge for $|\gamma| < 1.0745(2) \times 2 \pi$ 
and $|\gamma|^{-1} < [1.0745(2) \times  2 \pi]^{-1}$, respectively.
Our result for the convergence of the small $\gamma$
series is consistent with what is reported
in the literature~\cite{kvaal}.

{\em{Weakly-repulsive $(N,1)$ system, $N=2,3,\cdots$:}}
To treat the weakly-interacting system with finite $N$,
$N>1$, we rewrite the system Hamiltonian in 
second quantization and expand the field operators
for the majority particles and the impurity
in terms of single particle harmonic oscillator 
states (see, e.g., Ref.~\cite{bookLeggett}).
The interaction matrix elements can be evaluated analytically
and
the first-order perturbation theory treatment for positive 
$g$ yields
\begin{eqnarray}
\label{eq_B1}
B^{(1)}(N)= \frac{2 \sqrt{N} \Gamma ( 1/2+N)}{\pi^2 N!}.
\end{eqnarray}
The first-order energy shift may be interpreted as 
the leading-order mean-field shift.
We find $ \lim_{N \rightarrow \infty} B^{(1)}(N)=2/\pi^2$, which agrees
with the coefficient obtained
by expanding Eq.~(\ref{eq_mcguire}). 
The evaluation of the second-order energy shift involves 
the evaluation of 
infinite sums. 
We find, as expected, that these
sums converge. The reason is that the
one-dimensional delta-function interaction does not,
unlike two- or three-dimensional delta-function
interactions~\cite{Huang57,2DPP}, require any
regularization if used in standard perturbation theory approaches.
We did not find a compact analytical expression 
applicable to all
$N$
for the second-order energy shift. 
For $N=1$ and $2$, we have
$B^{(2)}(1)=-\ln(2)/\pi^3$ and
$B^{(2)}(2)=[-9+6\sqrt{3}+3 \ln(2+\sqrt{3})-12 \ln(2)]/(4 \pi^3)$.
For larger $N$, the expressions are lengthy. The numerical values for
$N \le 12$ are listed in Table~\ref{tab1}.
Table~\ref{tab1} also summarizes the numerically determined values for the
third-order coefficients $B^{(3)}(N)$ for $N=2-6$.
The $B^{(3)}(N)$ coefficient increases slightly as $N$ changes from $1$
to $2$, and then decreases monotonically as $N$ increases further.
The numerically determined $B^{(3)}(N)$ coefficients
for $N=2-6$ 
approach the $N = \infty$ coefficient smoothly if plotted as a 
function of $1/N$.

{\em{Strongly-interacting $(N,1)$ system, $N=2,3,\cdots$:}}
The strongly-interacting regime
has been treated perturbatively
at leading order, i.e., at order $1/\gamma$, 
for $N \le 8$~\cite{levinsen}
(note, though, that only the coefficients for $N \le 4$ were reported
explicitly, i.e., in equation or numerical form). 
To derive these results,
the two-body interaction
for large $|g|$ is modeled by
imposing the two-body boundary condition 
on the many-body wave function when the distance between the unlike particles
approaches zero~\cite{volosniev1,levinsen,deuretzbacher}.
Since the ground state eigenenergy for $|g|=\infty$ is degenerate,
the perturbation shift is obtained by diagonalizing the
perturbation matrix constructed using the degenerate states
for $g=\infty$.
For the many-body states $\Psi_{\alpha}$ and $\Psi_{\beta}$, 
the perturbation matrix element $V_{\alpha \beta}$
reads~\cite{volosniev1}
\begin{eqnarray}
\label{eq_PTmatrixelement}
V_{\alpha \beta} = 
-\frac{\hbar^4}{m^2 g} \times \nonumber \\
\sum_{j=1}^N \int \cdots \int
\left(
\frac{\partial \Psi_{\alpha}^*}{\partial z_{j0}}\bigg|_{z_{j0} \rightarrow 0^+} -
\frac{\partial \Psi_{\alpha}^*}{\partial z_{j0}}\bigg|_{z_{j0} \rightarrow 0^-} 
\right )
\delta(z_{j0}) \times \nonumber \\
\left(
\frac{\partial \Psi_{\beta}}{\partial z_{j0}}\bigg|_{z_{j0} \rightarrow 0^+} -
\frac{\partial \Psi_{\beta}}{\partial z_{j0}}\bigg|_{z_{j0} \rightarrow 0^-} 
\right )
\text{d}z_0 \text{d}z_1 \cdots \text{d}z_N.
\end{eqnarray}
These matrix elements are closely related to the 
boundary condition representation of the one-dimensional odd-parity
pseudo-potential~\cite{GirardeauOlshanii,CheonShigehara}.
We evaluate the integrals in Eq.~(\ref{eq_PTmatrixelement})
analytically 
for $N=1-4$.
The analytical results for $N=1$ and $2$ read
$C^{(1)}(1)=-2\sqrt{\pi}$ and $C^{(1)}(2)=-\sqrt{\pi/2}(81/32)$.
The analytical expressions for $N=3$ and
$4$ are lengthy and not reported here~\cite{footnotePRL}.
For larger $N$, we perform all but one integration for each
of the perturbation matrix elements analytically.
The resulting numerically determined energy shifts 
are accurate to 
more than 10 digits.
Table~\ref{tab2} summarizes the numerical
values for the coefficient $C^{(1)}(N)$  for $N\le 9$
obtained by us.
The extension to larger
$N$ is, although tedious, possible in principle.

To determine the energy shift proportional to $\gamma^{-2}$,
we use second-order perturbation theory. 
Reference~\cite{Sen} pointed out that
the second-order perturbation theory energy shift of the $(1,1)$
system diverges, thus requiring regularization.
Analogous divergencies arise in the perturbative treatment
of one-dimensional
single-component Fermi gases 
with generalized delta-function
interactions (see, e.g., Ref.~\cite{GirardeauOlshanii})
and that of one-dimensional Bose gases with effective range
dependent zero-range interactions.
In the following,
we discuss the impurity problem with $N=2$ and 3. To evaluate the
second-order energy shifts,
we need to know the complete set of eigenstates of the 
$(2,1)$ and $(3,1)$ systems with $|g|=\infty$.
For the $(2,1)$ system, we use the analytical wave functions from 
Ref.~\cite{VolosnievThesis} and evaluate the
integrals analytically.
For the $(3,1)$ system, we derive compact forms for the eigenstates 
using spherical coordinates and evaluate the 
relevant integrals
analytically.
We then
evaluate the second-order
perturbation theory sums numerically, 
imposing an energy cutoff on the relative energy 
of the intermediate (or virtual) states that are being summed over.
The second-order energy shift 
is found to contain powerlaw divergencies
in the energy cutoff.
These divergencies are canceled through the introduction
of a counterterm and
the constant (and physically meaningful) part
is extracted with high precision
by a regularization scheme
similar to that developed 
for harmonically trapped bosons~\cite{johnson}.
Table~\ref{tab2} reports the resulting second-order perturbation 
theory coefficients with errorbars.
Our perturbative coefficients are consistent with the 
coefficients obtained 
by fitting the $(2,1)$ and $(3,1)$ energies reported in 
Refs.~\cite{Gharashi13,GharashiNote}
to a polynomial in $\gamma^{-1}$.
For the $N=2$ and $3$ systems, we extend the above treatment to the
third order (see Table~\ref{tab2}).
These third-order calculations require the evaluation of matrix
elements $V_{\alpha \beta}$ between 
excited states. Since the third-order perturbation expression
is more involved than the second-order perturbation expression, 
our third-order result has a larger errorbar than our
second-order result~\cite{energycutoff}.

The calculations of the second- and third-order energy shifts can,
in principle, be 
extended to larger $N$. To do so, two challenges
need to be overcome. 
First, an efficient method to generate 
the complete set of eigenstates at $|g|=\infty$ has to be devised.
Second, an efficient scheme for evaluating the matrix
elements and infinite perturbation
theory sums has to be developed.
This is not pursued here.

\section{Fitting the $B^{(k)}(N)$ and $C^{(k)}(N)$ 
coefficients} 
\label{sec_fitting}

Tables~\ref{tab1}-\ref{tab2} suggest that
the coefficients $B^{(k)}(N)$ and $C^{(k)}(N)$ 
change, for fixed $k$, 
smoothly with
$N$. This motivates us to 
write 
\begin{eqnarray}
\label{eq_fit1}
B^{(k)}(N)=
\sum_{j=0}^{j_{\text{max}}} b_j^{(k)} 
\left(\frac{1}{N} \right)^j
\end{eqnarray}
and
\begin{eqnarray}
\label{eq_fit2}
C^{(k)}(N)=
\sum_{j=0}^{j_{\text{max}}} c_j^{(k)} 
\left(\frac{1}{N} \right)^j.
\end{eqnarray}
It should be noted that the
expressions~(\ref{eq_fit1})-(\ref{eq_fit2}) 
reduce to $b_0^{(k)}$ and
$c_0^{(k)}$,
respectively,
in the $N \rightarrow \infty$
limit.
In the following, the parameters $b_j^{(k)}$
and $c_j^{(k)}$ 
are obtained by fitting the coefficients $B^{(k)}(N)$ and $C^{(k)}(N)$
for fixed $k$.

We start with $B^{(2)}(N)$.
We fit Eq.~(\ref{eq_fit1}) to the $B^{(2)}(N)$ values for $N=1-80$
(the values for $N=1-12$ are reported in Table~\ref{tab2}), varying 
$j_{\text{max}}$ from $2-20$. We find that the most reliable fit
is obtained for $j_{\text{max}}=12-13$. In this case, 
the fitting parameter $b_0^{(2)}$
differs from $-1/(4 \pi^2)$ [the result obtained
by expanding Eq.~(\ref{eq_mcguire})] by less than $10^{-8}$.
This suggests that not only the $k=1$ coefficient (see discussion above)
but also the $k=2$ coefficient connects smoothly with the
infinite $N$ result.
\begin{table}
\caption{Fitting coefficients $b_{j}^{(k)}$ 
for $k=2$ and $3$.
For $k=2$ and 3, we used 
$j_{\text{max}}=6$ and $4$, respectively.
}
\label{tab4}
\centering
\begin{tabular}{r@{=}l|r@{.}l r@{.}l}
\multicolumn{2}{c|}{} &   \multicolumn{2}{c}{$k=2$} & \multicolumn{2}{c}{$k=3$} \\
\hline
$j$&$0$ & $-0$&$0253304$ & $ 1$&$71100 \times 10^{-3}$ \\
$j$&$1$ & $ 0$&$0019591$ & $ 2$&$23905 \times 10^{-4}$ \\
$j$&$2$ & $ 0$&$0033477$ & $ 6$&$59881 \times 10^{-6}$ \\
$j$&$3$ & $-0$&$0116972$ & $-3$&$67025 \times 10^{-4}$ \\
$j$&$4$ & $ 0$&$0379414$ & $ 2$&$64073 \times 10^{-4}$ \\
$j$&$5$ & $-0$&$0684440$ & \multicolumn{2}{c}{} \\ 
$j$&$6$ & $ 0$&$0486346$ & \multicolumn{2}{c}{} 
\end{tabular}
\end{table}
Table~\ref{tab4} reports the results of our fit
to the $B^{(2)}(N)$ coefficients with $N=1-80$
and $\infty$ by a polynomial with $j_{\text{max}}=6$.

As mentioned earlier, the $B^{(3)}(1)$ coefficient
is slightly smaller than the $B^{(3)}(2)$ coefficient.
The $B^{(3)}(N)$ coefficients for $N \ge 2$, however, decrease monotonically.
This motivates us to fit the 
$B^{(3)}(N)$ coefficients with $N=2-6$ and $\infty$ by a polynomial
with $j_{\text{max}}=4$.
The fit coefficients are reported in Table~\ref{tab4}.
It can be seen that the coefficient $b_0^{(3)}$ agrees with the coefficient
$B^{(3)}(\infty)$ reported in Table~\ref{tab1}.
We believe
that our fit provides an accurate description of
the $6<N< \infty$ coefficients.

\begin{figure}
\centering
\includegraphics[angle=0,width=0.35\textwidth]{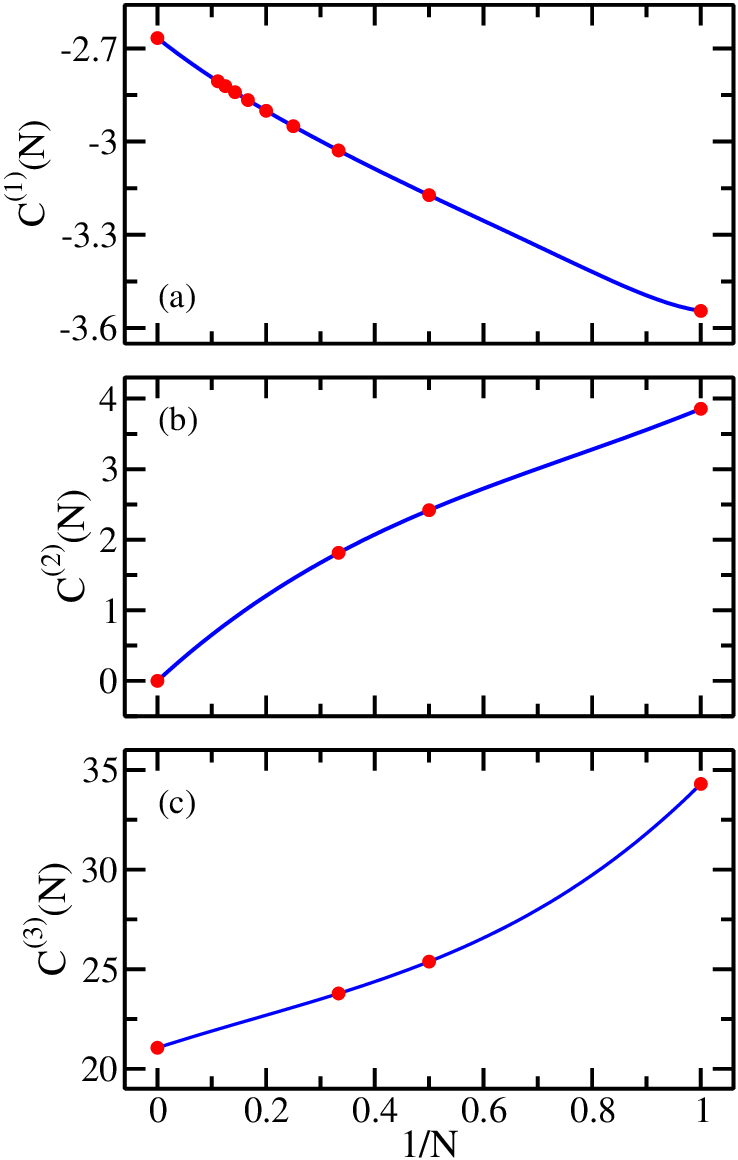}
\vspace*{0.5in}
\caption{(Color online)
Symbols show the coefficients 
(a) $C^{(1)}(N)$,
(b) $C^{(2)}(N)$, and
(c) $C^{(3)}(N)$
as a function of $1/N$.
The solid lines show our fits with $j_{\text{max}}=6$,
3, and 3, respectively.
}\label{fig3}
\end{figure}

Symbols in Figs.~\ref{fig3}(a)-\ref{fig3}(c)
show the $C^{(k)}(N)$ coefficients with $k=1, 2$, and $3$, respectively,
as a function
of $1/N$.
Our fits to these data (see Table~\ref{tab2})
using polynomials with $j_{\text{max}}=6,3$, and $3$
are shown by solid lines
(see Table~\ref{tab5} for the coefficients).
\begin{table}
\caption{Fitting coefficients $c_{j}^{(k)}$ 
for 
$k=1,2$, and $3$.
For $k=1,2$, and $3$, we used
$j_{\text{max}}=6,3$, and $3$, respectively.
}
\label{tab5}
\centering
\begin{tabular}{r@{=}l|r@{.}l r@{.}l  r@{.}l}
\multicolumn{2}{c|}{} &   \multicolumn{2}{c}{$k=1$} & \multicolumn{2}{c}{$k=2$} & \multicolumn{2}{c}{$k=3$}\\
\hline
$j$&$0$ & $-2$&$66667$ & $0$&$00000$ & $21$&$05520$ \\
$j$&$1$ & $-1$&$40749$ & $7$&$06739$ & $8$&$80915$ \\ 
$j$&$2$ & $ 1$&$78704$ & $-5$&$70589$ & $-5$&$07455$ \\
$j$&$3$ & $-4$&$21746$ & $2$&$49453$ & $9$&$51090$ \\
$j$&$4$ & $ 7$&$33094$ & \multicolumn{2}{c}{} & \multicolumn{2}{c}{} \\ 
$j$&$5$ & $-7$&$13604$ & \multicolumn{2}{c}{} & \multicolumn{2}{c}{} \\ 
$j$&$6$ & $ 2$&$76476$ & \multicolumn{2}{c}{} & \multicolumn{2}{c}{} 
\end{tabular}
\end{table}

The discussion so far has focused on the coefficients
$B^{(k)}(N)$  and $C^{(k)}(N)$ 
with $k=1-3$.
It is, in general, not feasible to extend the perturbative calculations
to higher $k$ for arbitrary $N$.
However, for $N=1$ and $\infty$, the coefficients with larger $k$
can be obtained readily.
We find that $|B^{(k)}(1)|$ decreases 
monotonically with increasing $k$ (we checked this for $k \le 50$).
The
$|C^{(k)}(1)|$ coefficient increases monotonically with increasing $k$
for $k < 37$; for $k \ge 37$, we observe small non-monotonic oscillations.
For $N = \infty$, we find that the $B^{(k)}(\infty)$ with 
$k$ even and $k \ge 4$
vanish while the $|B^{(k)}(\infty)|$ with $k$ odd decrease monotonically
with increasing $k$ (again, we checked this for $k \le 50$).
Similarly, the $C^{(k)}(\infty)$ with $k$ even and $k \ge 2$
vanish while the $|C^{(k)}(\infty)|$ with $k$ 
odd increase monotonically with increasing $k$.
Assuming a linear change with $1/N$, interpolating between $B^{(k)}(1)$
and $B^{(k)}(\infty)$ 
and 
between $C^{(k)}(1)$
and $C^{(k)}(\infty)$ 
for $k>3$ 
yields estimates for the finite $N$, $N>1$, coefficients.
While rough, these estimates might provide a reasonable
means to connect the weak and strong
perturbation theory limits for quantities such as those shown in
Figs.~\ref{fig4} and \ref{fig6}.

We cannot accurately estimate the radius of convergence of the small
and large $\gamma$ expansions for $1<N<\infty$.
However,
the fact that the radius of convergence is given by 
$|\gamma| < 1.0745(2) \times 2 \pi$ for $N=1$
and $\gamma <2 \pi$ for $N=\infty$
for the small $\gamma$ series
and by $|\gamma|^{-1} < [1.0745(2) \times 2 \pi]^{-1}$  for $N=1$
and
$\gamma^{-1} < (2 \pi)^{-1}$  for $N=\infty$
for the large $\gamma$ series suggests two speculations:
First, 
a convergent series can be found for any $\gamma$ and $N$.
Second, the radius of convergence of the small $\gamma$ series is
approximately $2 \pi$ for all $N$.
Figures~\ref{fig4} and \ref{fig6}, which are discussed
in the next section,
are consistent with these speculations.

\section{Discussion}
\label{sec_implication}

This section compares the perturbative energy expressions 
with the numerically determined energies of the upper branch.
Figure~\ref{fig1}(b) shows that the scaled interaction energy
$\epsilon(N)/E_{\text{F}}(N)$ depends weakly on $N$
if plotted as a function of 
$-\hbar \omega a_{\text{ho}}/g$. The dependence on $N$ is even weaker 
when
the interaction strength is parameterized by $\gamma$ as opposed 
to $g$.
To benchmark the applicability of the perturbative expressions
we analyze the interaction energy 
of the system with $N$ majority atoms
by comparing with that of the $(1,1)$ system.
Specifically, we consider the quantities $\rho(N)$,
\begin{eqnarray}
\label{eq_rho}
\rho(N)= \frac{\epsilon(N)/E_{\text{F}}(N)}{\epsilon(1)/E_{\text{F}}(1)},
\end{eqnarray}
and
$\delta(N,N')$, 
\begin{eqnarray}
\label{eq_delta}
\delta(N,N')=
\frac{\epsilon(N)/E_{\text{F}}(N)-\epsilon(1)/E_{\text{F}}(1)}
{\epsilon(N')/E_{\text{F}}(N')-\epsilon(1)/E_{\text{F}}(1)}.
\end{eqnarray}
For finite $N$, 
$\rho(N)$
reduces to
$\epsilon(N)/[N\epsilon(1)]$, i.e., $\rho(N)$ tells one 
the interaction energy per particle, normalized by the interaction energy
of the $(1,1)$ system.
The quantity $\delta(N,N')$ can alternatively be written
as $[\rho(N)-1]/[\rho(N')-1]$.

\begin{figure}
\centering
\includegraphics[angle=0,width=0.35\textwidth]{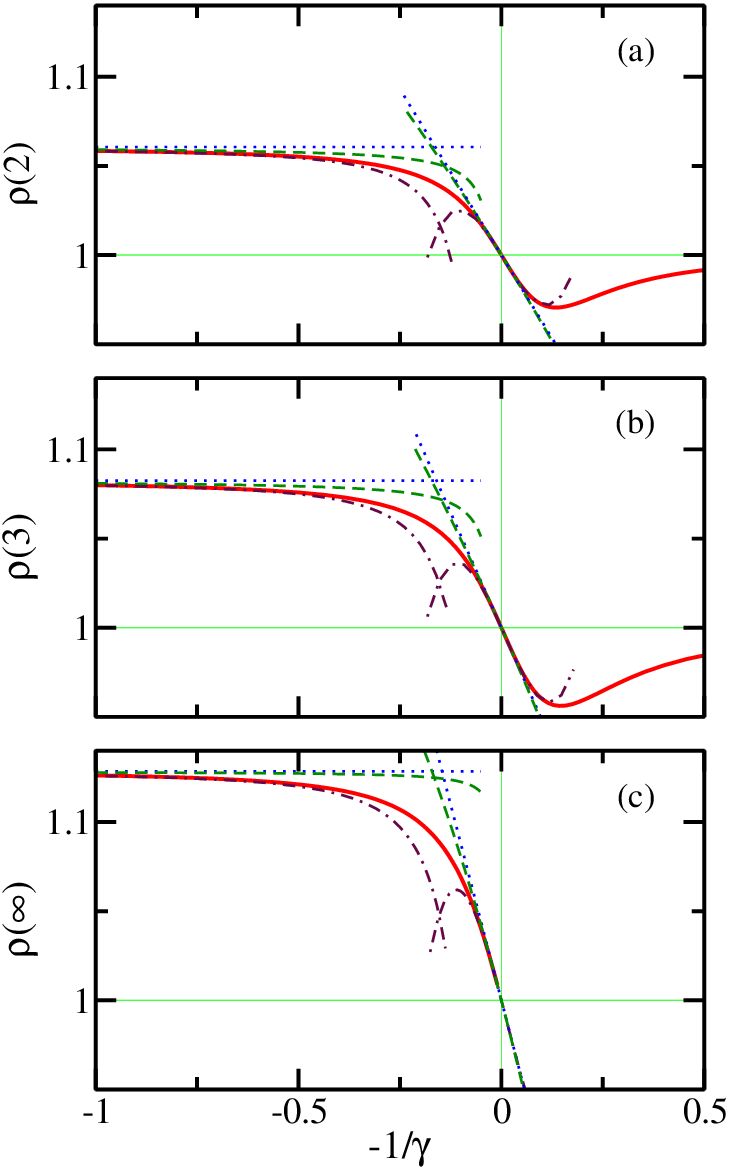}
\vspace*{0.5in}
\caption{(Color online)
Solid lines show the
quantity $\rho(N)$
as a function of $-1/\gamma$ for 
(a) $N=2$,
(b) $N=3$,
and 
(c) $N=\infty$, respectively.
For comparison, dotted, dashed, and dash-dotted lines show the
perturbative results for $\rho(N)$ accounting for terms up to
order $\gamma^0$, $\gamma^1$ and $\gamma^2$, respectively,
in the weakly-interacting regime and 
accounting for terms up to 
order $\gamma^{-1}$, $\gamma^{-2}$ and $\gamma^{-3}$, respectively,
in the strongly-interacting regime.
}\label{fig4}
\end{figure}

Expanding Eq.~(\ref{eq_rho})
in the weakly-interacting (small $\gamma$) regime, 
we find
\begin{eqnarray}
\rho(N) = \rho_0^{(\text{w})}(N) +
\rho_1^{(\text{w})}(N) \gamma +
\rho_2^{(\text{w})}(N) \gamma^2 + {\cal{O}}(\gamma^3),
\end{eqnarray}
where
the coefficients $\rho_k^{(\text{w})}(N)$ are determined by the 
$B^{(l)}(N)$
and $B^{(l)}(1)$ with $l \le k+1$. 
Expanding Eq.~(\ref{eq_rho})
in the strongly-interacting (large $|\gamma|$) regime, 
we find
\begin{eqnarray}
\rho(N) = 1 + \rho_1^{(\text{s})}(N) \gamma^{-1} +
\rho_2^{(\text{s})}(N) \gamma^{-2} + \nonumber \\
\rho_3^{(\text{s})}(N) \gamma^{-3} + {\cal{O}}(\gamma^{-4}),
\end{eqnarray}
where
the coefficients $\rho_k^{(\text{s})}(N)$ are determined by the 
$C^{(l)}(N)$
and $C^{(l)}(1)$ with $l \le k$. 
Solid lines in Figs.~\ref{fig4}(a)-\ref{fig4}(c) show the 
quantity
$\rho(N)$
for $N=2$, 3, and $\infty$, respectively.
The solid lines are obtained using the numerical
$(2,1)$ and $(3,1)$ energies and the semi-analytical $(1,1)$
and $(\infty,1)$ energies.
For $\gamma \rightarrow 0^+$ (i.e., 
for $-1/\gamma \rightarrow -\infty$),
the quantity $\rho(N)$ approaches the constant
$\rho_0^{(\text{w})}(N)=B^{(1)}(N)/B^{(1)}(1)$
[see the horizontal
dotted lines in Figs.~\ref{fig4}(a)-\ref{fig4}(c)], which 
increases monotonically 
from
$1.0607$ to $1.1284$ as $N$ goes from $2$ to $\infty$.
This portion of the interaction energy can be interpreted as
the mean-field contribution.
Inclusion of the next order correction 
[the $\rho_1^{(\text{w})}(N) \gamma$ term]
and the next two
corrections 
[the $\rho_1^{(\text{w})}(N) \gamma$ and
$\rho_2^{(\text{w})}(N) \gamma^2$ 
terms]
yields the dashed and dash-dotted lines in 
Figs.~\ref{fig4}(a)-\ref{fig4}(c).
The dash-dotted lines
provide a fairly accurate description of the
quantity $\rho(N)$ for $-1/\gamma \lesssim -0.4$. 
For $|\gamma| \rightarrow \infty$, the 
leading-order $\gamma$-dependent term [see the
(non-horizontal) dotted 
lines in Figs.~\ref{fig4}(a)-\ref{fig4}(c)]
increases monotonically 
from 0.3725 to 0.8782 as $N$ changes from $2$ to $\infty$.
Inclusion of the next-order correction and the next two
corrections yields the
dashed and dash-dotted lines in Figs.~\ref{fig4}(a)-\ref{fig4}(c).
It can be seen that the dash-dotted lines
provide a fairly accurate
description of the quantity $\rho(N)$ for $-1/\gamma \gtrsim -0.15$.
This value is close to the expected radius of convergence
of the interaction energy
[recall, 
the radius of convergence
is $1/\gamma=(1.0745 \times 2 \pi)^{-1} \approx 0.148$
for the $(1,1)$ system].
Combining the perturbative descriptions for small and large $|\gamma|$,
the expansions provide a fairly accurate description of the
interaction energy for the system with $N$ majority particles,
normalized by that for the $(1,1)$ system, 
over a wide range of interaction strengths $\gamma$.

\begin{figure}
\centering
\includegraphics[angle=0,width=0.35\textwidth]{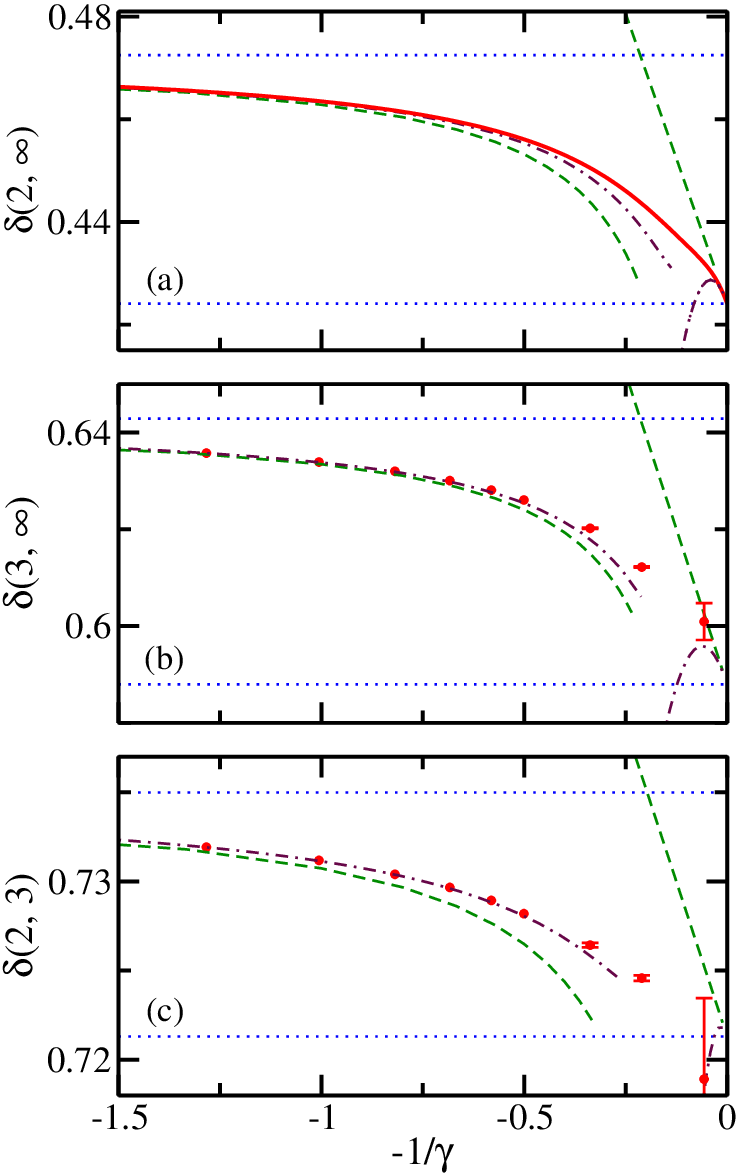}
\vspace*{0.5in}
\caption{(Color online)
The
quantity $\delta(N,N')$
as a function of $-1/\gamma$.
The solid line is for
(a) $N=2$ and $N'=\infty$ and
the circles are for
(b) $N=3$ and $N'=\infty$
and 
(c) $N=2$ and $N'=3$.
In the large $\gamma$ regime, the uncertainty of the
numerically determined $(3,1)$ energies leads to 
appreciable uncertainties in 
$\delta(2,3)$ and $\delta(3,\infty)$ [see the errorbars in
Figs.~\ref{fig6}(b) and \ref{fig6}(c)].
For comparison, dotted, dashed, and dash-dotted lines show the
perturbative results for $\delta(N,N')$ accounting for terms up to
order $\gamma^0$, $\gamma^1$, and $\gamma^2$, respectively,
in the weakly-interacting, small $|\gamma|$ regime and 
accounting for terms up to 
order $\gamma^{-1}$, $\gamma^{-2}$, and $\gamma^{-3}$, respectively,
in the strongly-interacting, large $|\gamma|$ regime.
}\label{fig6}
\end{figure}

Expanding Eq.~(\ref{eq_delta})
in the weakly-interacting regime, 
we find
\begin{eqnarray}
\delta(N,N') = 
\delta_0^{(\text{w})}(N,N') +
\delta_1^{(\text{w})}(N,N') \gamma + \nonumber \\
\delta_2^{(\text{w})}(N,N') \gamma^2 + {\cal{O}}(\gamma^3),
\end{eqnarray}
where
the coefficients $\delta_k^{(\text{w})}(N,N')$ are determined by the 
$B^{(l)}(N)$, $B^{(l)}(N')$,
and $B^{(l)}(1)$ with $l \le k+1$. 
Expanding Eq.~(\ref{eq_delta})
in the strongly-interacting regime, 
we find
\begin{eqnarray}
\delta(N,N') = 
\delta_0^{(\text{s})}(N,N')  +
\delta_1^{(\text{s})}(N,N') \gamma^{-1} + \nonumber \\
\delta_2^{(\text{s})}(N,N') \gamma^{-2} + {\cal{O}}(\gamma^{-3}),
\end{eqnarray}
where
the coefficients $\delta_k^{(\text{s})}(N,N')$ are determined by the 
$C^{(l)}(N)$, $C^{(l)}(N')$,
and $C^{(l)}(1)$ with $l \le k+1$. 
The 
quantity
$\delta(N,N')$
is shown by the solid line in Fig.~\ref{fig6}(a)
for $(N,N')=(2,\infty)$
and by dots in Figs.~\ref{fig6}(b)-\ref{fig6}(c) 
for
$(3,\infty)$ and $(2,3)$, respectively.
We observe that the quantity $\delta(N,N')$ changes 
only slightly as $-1/\gamma$ goes from $-\infty$ to $0$;
this is particularly true for 
$\delta(2,3)$ [see Fig.~\ref{fig6}(c)].
The limiting values 
[see the dotted lines in Figs.~\ref{fig6}(a)-\ref{fig6}(c)]
are given by $\delta_0^{(\text{w})}(N,N')$ 
and $\delta_0^{(\text{s})}(N,N')$,
respectively.
Dashed lines include
the next order correction in the weakly- and strongly-interacting
regimes, and dash-dotted lines include the
next two corrections.
In the weakly-interacting regime,
the dash-dotted lines provide a fairly good description of the
quantity $\delta(N,N')$.
In the strongly-interacting regime, however, the 
validity regime of the perturbative expressions is quite small.
For $\delta(2,3)$, e.g., the expansion coefficients are
$\delta_0^{(\text{s})}(2,3)=0.7213$,
$\delta_1^{(\text{s})}(2,3)=0.0694(3)$,
and
$\delta_2^{(\text{s})}(2,3)=-2.31(12)$,
where the numbers in brackets denote the errorbars
due to the uncertainties of the second- and third-order
perturbation theory coefficients. 
The fact that 
$|\delta_2^{(\text{s})}(2,3)| \gg |\delta_1^{(\text{s})}(2,3)|$
is responsible for the turn-around of the dash-dotted line for large 
positive $\gamma$.
We note that the errorbar of the quantities
$\delta(3,\infty)$ and $\delta(2,3)$,
obtained from the numerical energies [see dots in
Figs.~\ref{fig6}(b)-\ref{fig6}(c)], is too large 
in the
large $\gamma$ regime to meaningfully 
compare with the perturbative results.

\section{Conclusion}
\label{sec_conclusion}

This paper considered the upper branch of 
a non-interacting harmonically trapped
one-dimensional Fermi gas with a single impurity.
Zero-range two-body contact interactions 
with strength $g$ were assumed between
the majority atoms and the impurity.
This system constitutes one of the simplest mesoscopic systems
accessible to experiment and theory.
On the experimental side, it has been demonstrated by the Heidelberg group
that the upper branch of the model Hamiltonian can be emulated
reliably using ultracold atoms~\cite{Jochim1,Jochim3}.
On the theory side, various numerical and 
analytical techniques have been applied~\cite{astrakharchik,mueller,brouzos,harshman,Gharashi12,Gharashi13,conduit,lewenstein,volosniev1,volosniev2,ho,levinsen}.
This paper pursued a perturbative approach,
which determined expansions of the energy of the upper branch
in the weakly- and strongly-interacting regimes for various
$N$. In the cases where we were not able
to obtain general $N$ expressions
for a fixed order in the perturbative expansion,
approximate expressions applicable to all $N$ were obtained 
through fits.
Through comparison with accurate numerical
few-body energies,
the perturbative expressions were shown to
provide a satisfactory description for a wide range
of interaction strengths.

The main results of this work are:
(i) 
We determined an expansion for the energy of the upper branch 
of a one-dimensional harmonically trapped
Fermi gas with a single 
impurity in the weakly-repulsive regime up to order 
$\gamma^3$, applicable to any system size.
(ii)
We determined an expansion for the energy of the upper branch 
in the strongly-interacting regime up to order 
$\gamma^{-3}$, applicable to any system size.
While the idea to treat the coupling strength $1/\gamma$
as a small parameter is not new, our work provides an explicit
demonstration that such a program can be carried through explicitly
beyond the leading-order correction.
(iii)
The radii of convergence of the series were reported for $N=1$ and $\infty$.
(iv)
The behavior of the expansion coefficients
in the series in $\gamma^k$ and $\gamma^{-k}$ with $k>3$
was discussed.
(v)
The perturbative expressions were benchmarked and 
found to provide a reliable description over a wide
range of interaction strengths.

The results presented in this work can be used to calculate perturbative
expressions for the contact and other observables.
Moreover, the second- and third-order results
in the $\gamma^{-1}$ series allow one to assess the applicability
regime of effective spin models~\cite{levinsen,deuretzbacher,zinnerspin}.

Acknowledgement:
We thank G. Z\"urn for email correspondence that motivated this work
and suggested to analyze the quantity
$\delta(N,N')$ shown in Fig.~\ref{fig6}.
We 
also thank N. Zinner for bringing Ref.~\cite{Sen} to our attention,
and P. Johnson and E. Tiesinga for helpful discussions on the numerical
regularization of diverging sums.
Support by the National Science Foundation  through grant number
PHY-1205443 is
 gratefully acknowledged.



\begin{thebibliography}{10}
\bibitem{GharashiNote}
In the vicinity of $|g| \rightarrow \infty$, we improved 
the numerical energies reported in Ref.~\cite{Gharashi13}.
\bibitem{series}
Throughout, we refer to the expansion around $\gamma=0^+$ 
and $|\gamma^{-1}|=0$ as small and large $\gamma$ series.
Sections~\ref{sec_pert} and 
\ref{sec_fitting} discuss the applicability regime, i.e., the
radius of convergence, of these series.
\end{thebibliography}
\end{document}